%% file: hadron2011.tex
\documentclass[a4paper,11pt]{article}

\usepackage{contribution}

\input{econfmacros}
\input{contributionmacros}

\begin{document}

\input{contribution}

\end{document}

%% file: econfmacros.tex
\newcommand{\weblink}[2][]{%
    \ifthenelse{\equal{#1}{}}%
    {\textnormal{\url{#2}}}%
    {\textnormal{\href{#2}{#1}}}%
}

\newcommand{\acknowledgements}[1]{%
  \bigskip\bigskip
  \textsf{\textbf{\Large Acknowledgements}} \\[2ex]
  {#1}
  \bigskip
}

\def\beq{\begin{equation}}
\def\eeq#1{\label{#1}\end{equation}}
\def\eeqn{\end{equation}}

\def\beqa{\begin{eqnarray}}
\def\eeqa#1{\label{#1}\end{eqnarray}}
\def\eeqan{\end{eqnarray}}

\let\bar=\overbar

\def\etal{{\it et al.}}

\def\Dslash{\not{\hbox{\kern-4pt $D$}}}
\def\dslash{\not{\hbox{\kern-2pt $\del$}}}

\def\msb{{\bar{\ssstyle M \kern -1pt S}}}

%% file: contributionmacros.tex
\newcommand{\contribution}[7][]{%
  \clearpage
  \thispagestyle{plain}
  \ifthenelse{\equal{#1}{}}
  {\hypersetup{pdftitle={#2}}}
  {\hypersetup{pdftitle={#1}}}
  \hypersetup{pdfauthor={{#3} {#4}}}
  {\centering\normalfont\LARGE\bfseries\sffamily #2 \par\nobreak}
  \lhead{}
  \chead{%
    \textit{\footnotesize XIV International Conference on Hadron Spectroscopy
      (\weblink[\textit{hadron2011}]{http://www.hadron2011.de}), 13-17 June 2011, Munich, Germany}%
  }
  \rhead{}
  \bigskip
  \begin{center}
    {#3} {#4}\ifthenelse{\equal{#6}{}}{}{\footnote{\weblink[#6]{mailto:#6}}}
    \ifthenelse{\equal{#7}{}}{}{#7} \\
    \textit{#5}
  \end{center}
  \bigskip
}

\renewcommand{\abstract}[1]{%
  \begin{center}
    \begin{minipage}{0.85\textwidth}
      \begin{footnotesize}
        #1
      \end{footnotesize}
    \end{minipage}
  \end{center}
  \bigskip
}

%% file: contribution.tex
%
%
%
%
%
{  


%

\contribution[First Results from VEPP2000 Collider]  
{First results from the CMD3 Detector  \\ at the VEPP2000 Collider}  
{Evgeny P.}{Solodov}  
{Budker Institute of Nuclear Physics \\
  Novosibirsk, Russia }  
{solodov@inp.nsk.su}  
{on behalf of the CMD3 Collaboration}  
%

\abstract{%
Regular data taking started at the VEPP2000 $e^+ e^-$  
  Collider with  CMD3 and SND detectors. Energy scan
  for center-of-mass energy from 1 GeV to 2 GeV has been performed with about
  20 pb$^{-1}$ per detector. We present first preliminary results from
  the CMD3 detector. 
}
%

\section{Introduction}

Production of low energy hadrons in  $e^+ e^-$ collisions remains
an interesting experimental area due to its important contribution to
the Standard Model (SM) calculations of the muon anomalious magnetic moment
and $\alpha(s)$. 

   
An $e^+ e^-$ collider of the next generation, VEPP2000~\cite{vepp}, has been
constructed and started
regular data taking in BudkerINP, Novosibirsk, Russia. It is designed
to cover a center-of-mass ($E_{c.m.}$) energy from hadron production
threshold up to 2 GeV. 

Two detectors~\cite{sndcmd3}, SND and CMD3, have been
prepared for the rich physics program at the VEPP2000 collider.  
During next few years we plan to scan the available energy range to
measure the hadron production cross sections with a percent or better
accuracy level, as well as a study production dynamics for the multi-hadron
channels. 

In this paper we present preliminary results from the first energy
scan of the 1-2 GeV
center-of-mass energy region obtained with the CMD3 detector.

\section{The VEPP2000 Collider}

The VEPP2000 collider is described elsewere~\cite{vepp} and the layout is shown
in Fig.~\ref{vepplayout}. A special feature of the machine is the using of
the solenoidal focusing for the interaction regions. This new approach
allows to suppress beam-beam effects and store larger currents.
During the energy scan, reported here, a luminosity up
to 2$\cdot$10$^{31}$~cm$^{-2}$sec$^{-1}$ has been demonstrated,
limited by the positron current.  With a new positron source, currently under
construction, the designed luminosity is 10$^{32}$
cm$^{-2}$sec$^{-1}$.

\begin{figure}[htb]
  \begin{center}
    \includegraphics[width=0.7\textwidth]{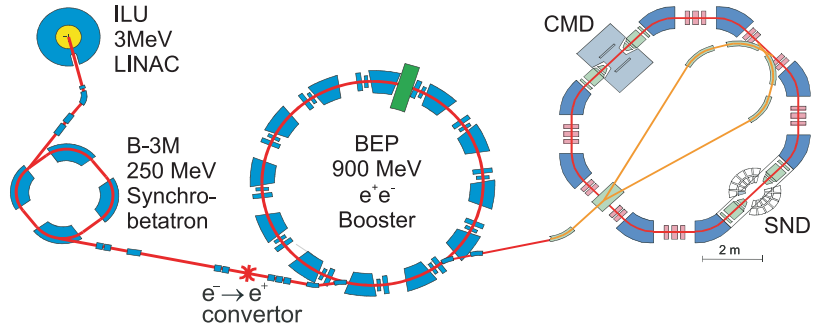}
    \caption{The layout of the VEPP2000 complex. The locations of the
    CMD3 and SND detectors are shown.}
    \label{vepplayout}
  \end{center}
\end{figure}
%
\begin{figure}[htb]
  \begin{center}
    \includegraphics[width=0.5\textwidth]{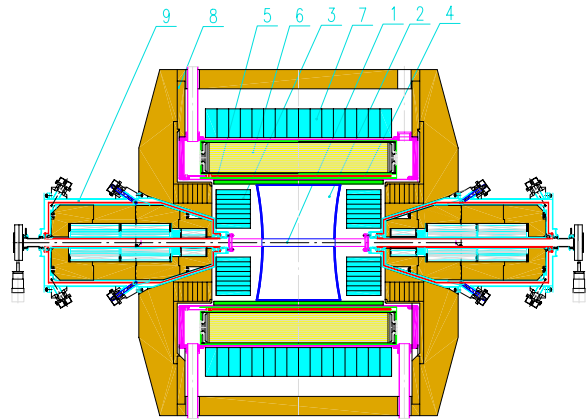}
    \caption{The CMD3 detector: 1-Interaction region; 2-Drift chamber;
    3-BGO end cap calorimeter; 4-Z-proportional chamber; 5-SC magnet  ;
    6-LXe calorimeter; 7-CsI calorimeter; 8-Yoke; 9-Focusing solenoids.}
    \label{cmdlayout}
  \end{center}
\end{figure}
%

\section{The CMD3 Detector}

The CMD3 detector is described elsewere~\cite{sndcmd3} and detector
elements are shown in Fig.~\ref{cmdlayout}. It is a general purpose
magnetic detector, providing good spatial and momentum resolutions for the charged
particles~\cite{dc}, and very good (about 1-2 mm) spatial resolution for photons
in the LXe calorimeter~\cite{lxe}, as well as good photon energy measurement.
The detector performance is demonstrated in Fig.~\ref{cmdcoll}, where
DC and calorimeter responses are shown for collinear events at
E$_{c.m.}$=1.975 GeV.
\begin{figure}[htb]
  \begin{center}
    \includegraphics[width=0.4\textwidth]{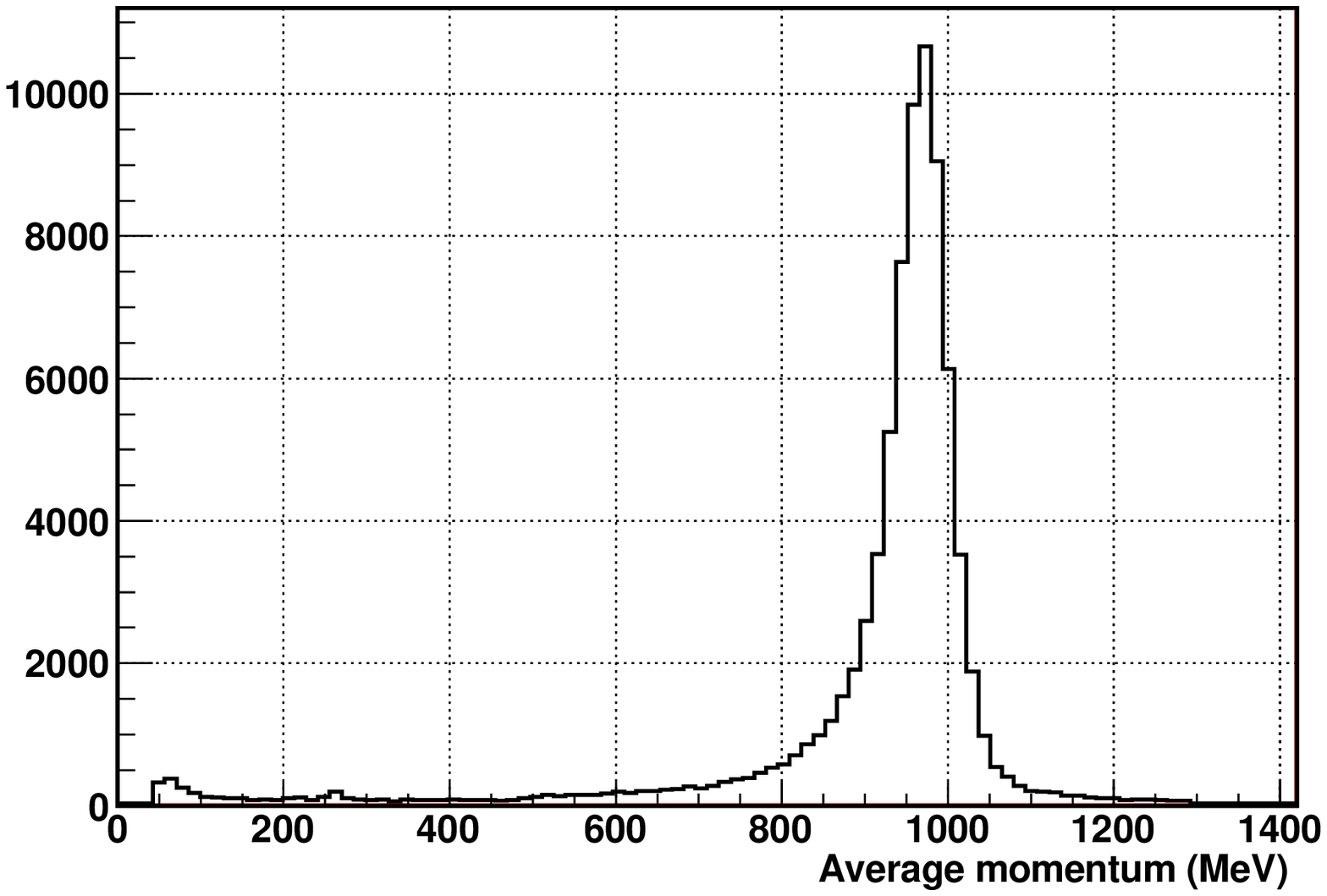}
    \includegraphics[width=0.4\textwidth]{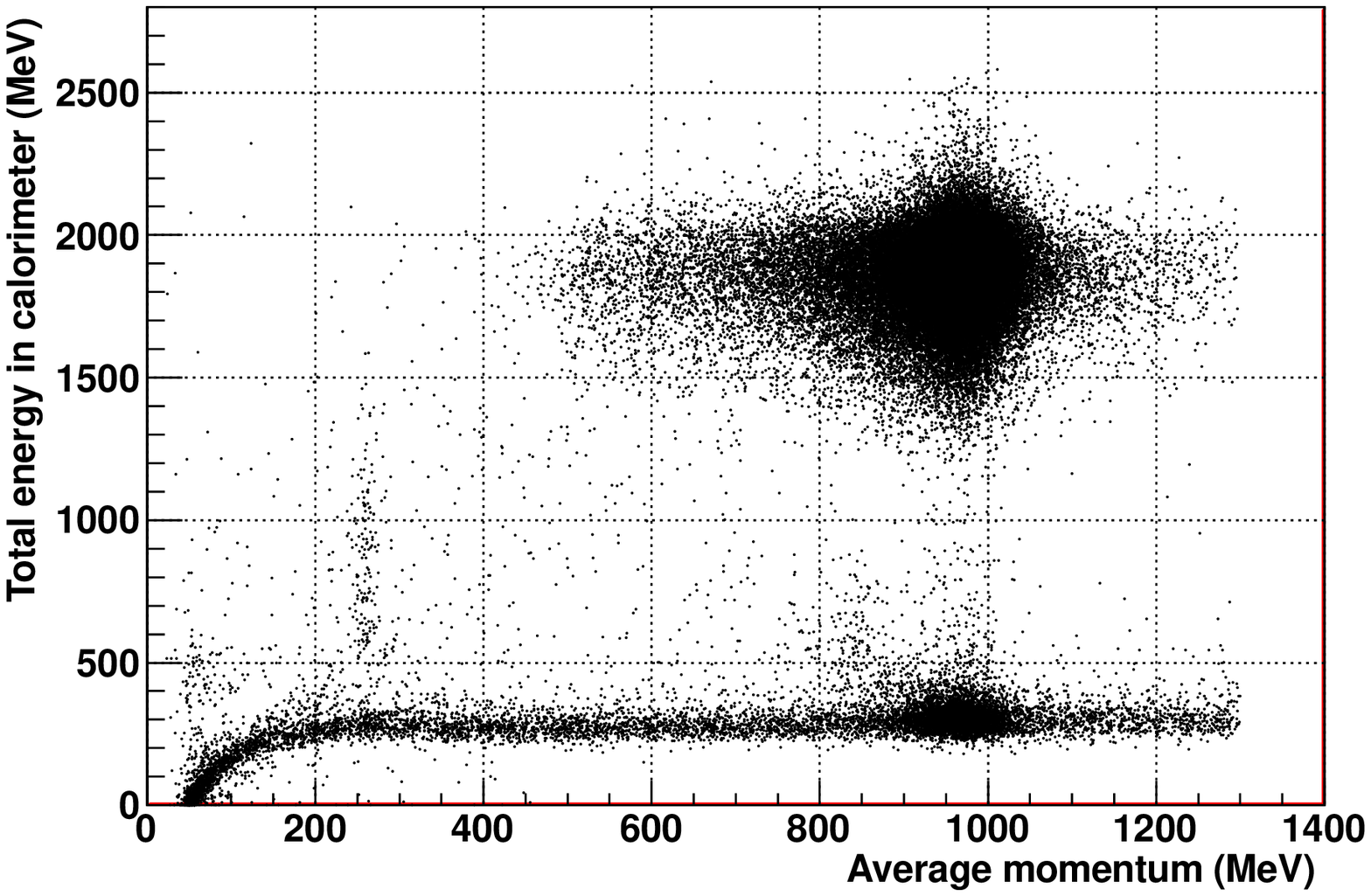}
    \includegraphics[width=0.4\textwidth]{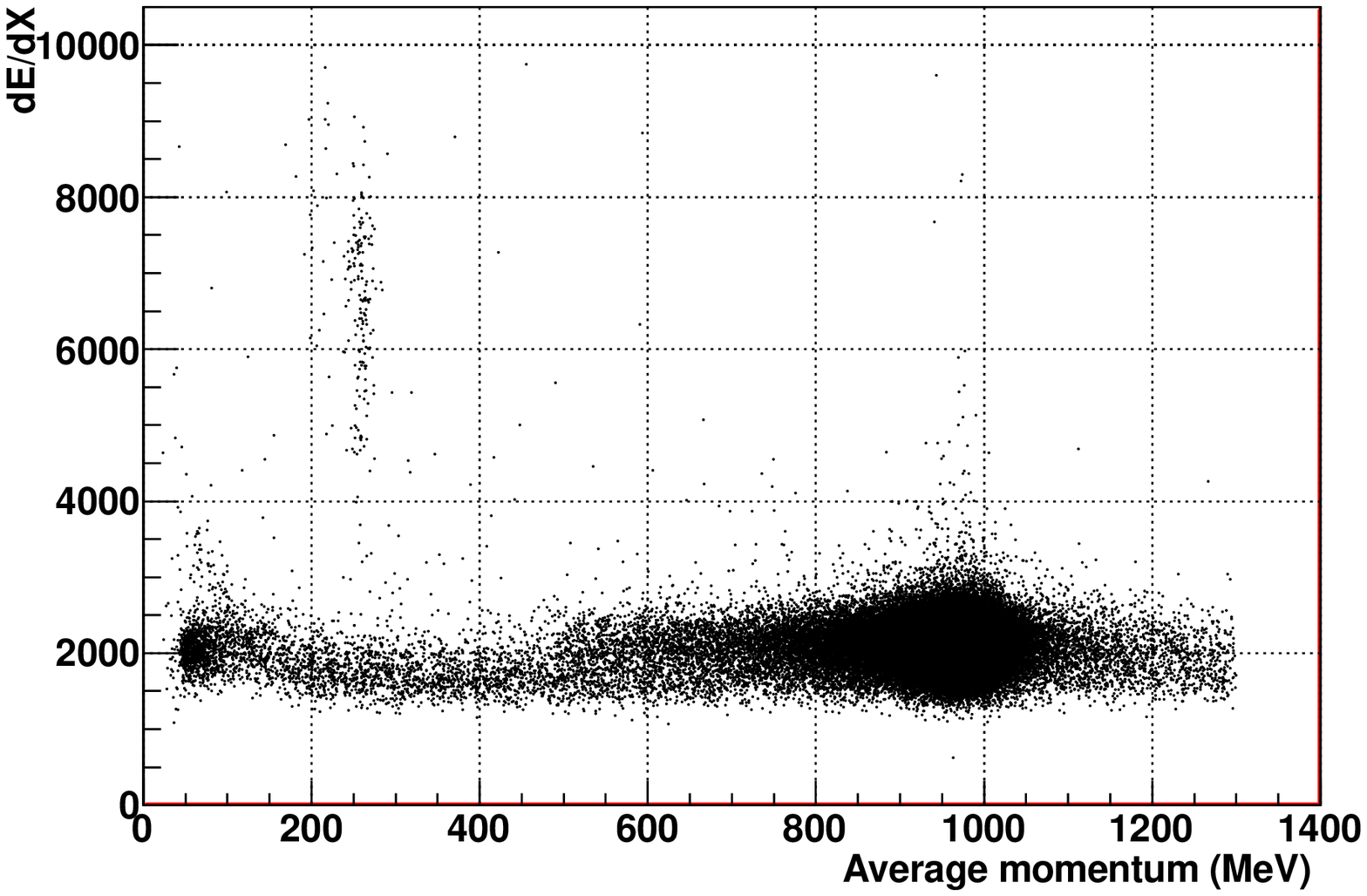}
    \includegraphics[width=0.4\textwidth]{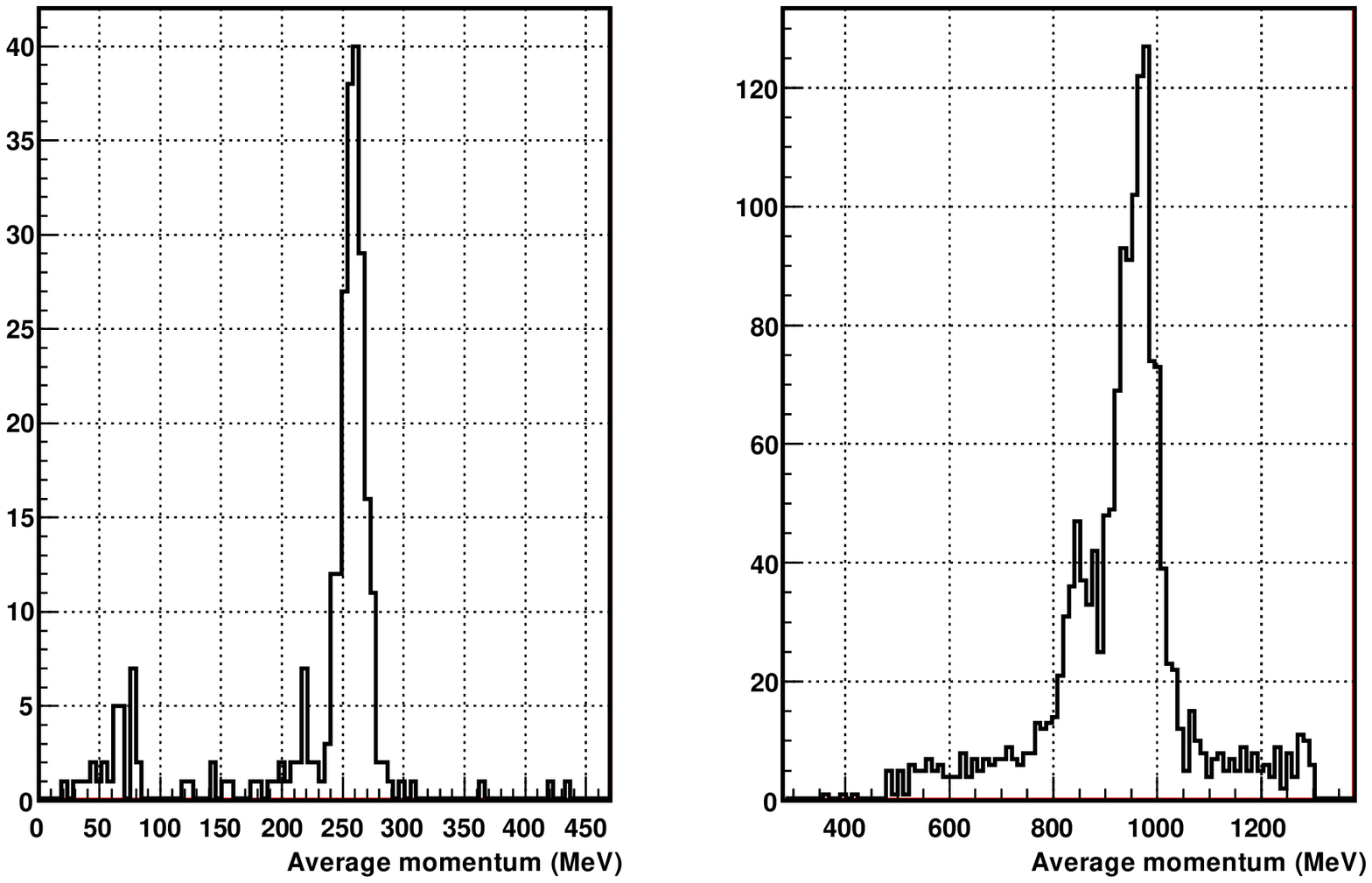}
    \caption{The CMD3 detector performance at E$_{c.m}$=1.975 GeV for
      collinear events:
      (top left) the average momentum ; (bottom left) the dedx DC
      measurement versus average momentum; (top right) the total
      energy deposition in the calorimeter versus average momentum;
      (bottom right) the $e^+ e^-\to P\bar P$ signal from dedx$>$3000
      selection and $e^+ e^-\to K^+ K^-$ and $e^+ e^-\to \pi^+ \pi^-$
      signals from calorimeter energy deposition selection.
    }
    \label{cmdcoll}
  \end{center}
\end{figure}
%
A relatively clean selection of the processes  $e^+ e^-\to e^+ e^-$, $P\bar P$, $
K^+ K^-$, $\pi^+ \pi^-$ can be performed using detector subsystems.

\section{First Physics results}

We perform the energy scan in the 1-2 GeV center-of-mass energy,
collecting data at 40 energy points with about 0.5 pb$^{-1}$
integrated luminosity. This luminosity corresponds to 200000 to
50000 events of Bhabha events (per point) used for the luminosity measurements and  
from a few hundred to a few thousand of multihadrons events like
$\pi^+\pi^-\pi^0$, $2(\pi^+\pi^-)$, $2(\pi^+\pi^-)\pi^0$, $K^+
K^-\pi^+\pi^-$, $6\pi$ etc.

As shown in Fig.~\ref{cmdcoll}(bottom), a simple requirement of
dEdX$>$3000 gives a very clean signal of the $e^+ e^-\to P\bar P$
process with about 200 events per energy point. We estimate the cross
section for four energy points 
above the threshold and  show in Fig.~\ref{cmdpbarp} a comparison of
our preliminary results with other measurements.  
\begin{figure}[htb]
  \begin{center}
    \includegraphics[width=0.99\textwidth]{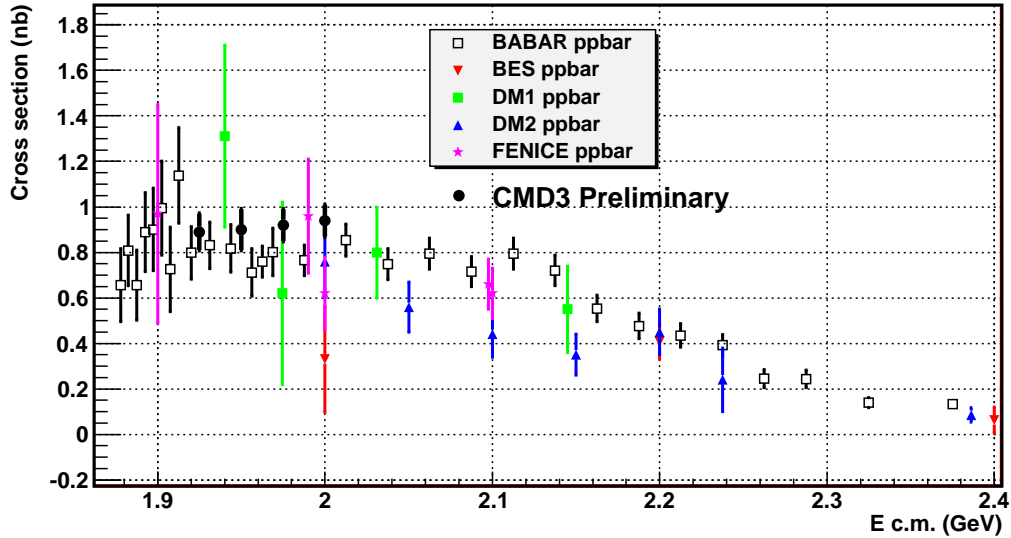}
    \caption{The CMD3 study of the $e^+ e^-\to P\bar P$ process in
    comparison with other measurements.
    }
    \label{cmdpbarp}
  \end{center}
\end{figure}
%

In this paper we also show our preliminary measurement of the $e^+
e^-\to 3(\pi^+\pi^-)$ cross section. We detect five and six charged
tracks and using the total energy for six tracks and missing mass for five
tracks select candidates for the $e^+ e^-\to 3(\pi^+\pi^-)$
reaction with almost no background, as shown in
Fig.~\ref{cmd6pi}(left). Our cross section measurement is shown in
Fig.~\ref{cmd6pi}(right) in comparison with recent BaBar data.

\begin{figure}[htb]
  \begin{center}
    \includegraphics[width=0.38\textwidth]{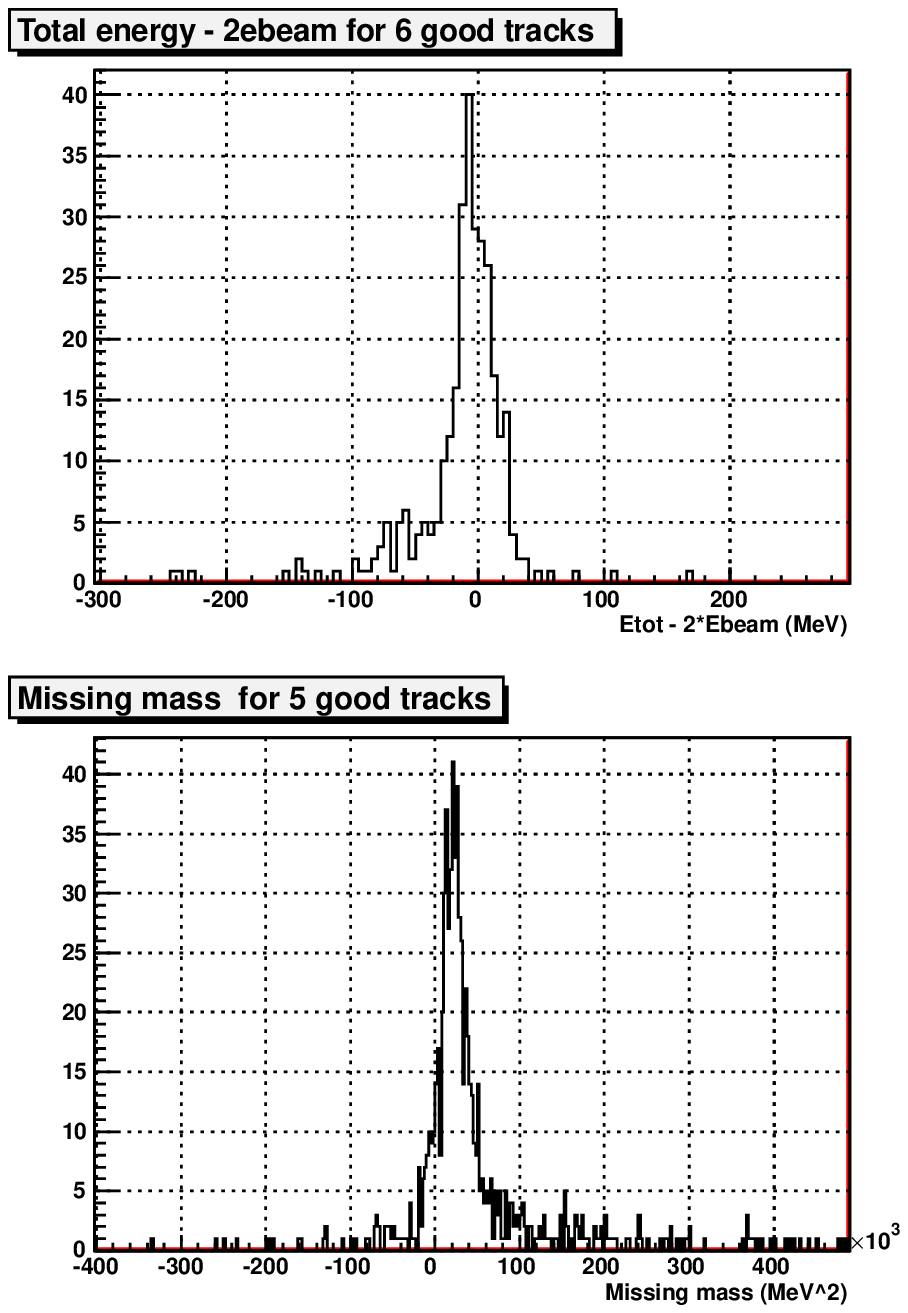}
    \includegraphics[width=0.52\textwidth]{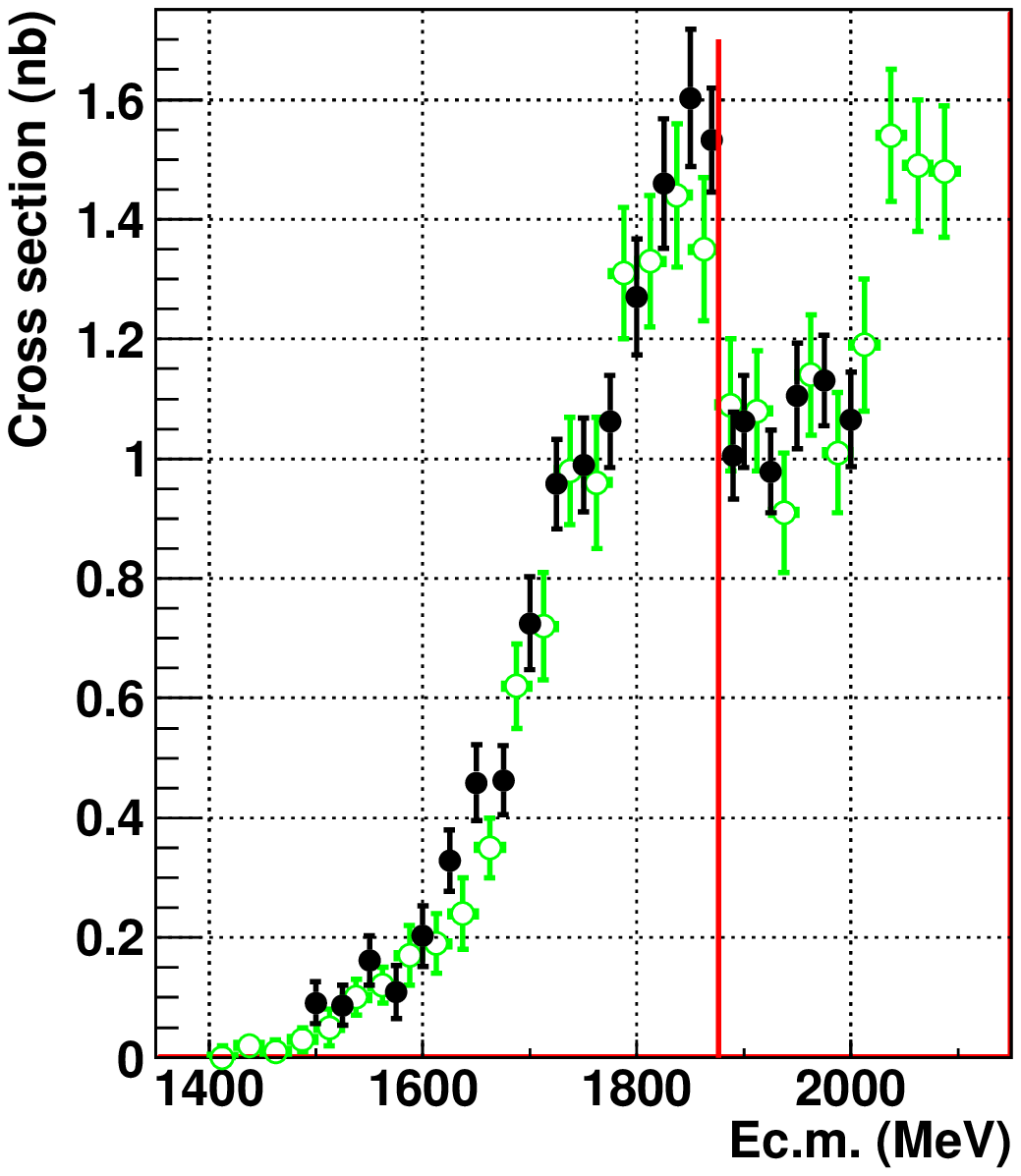}
    \caption{The CMD3 study of the $e^+ e^-\to 3(\pi^+\pi^-)$ process:
(left) the difference of the 6 pions total energy and E$_{c.m.}$ (top)
      and missing mass for 5 tracks (bottom);
      (right) the $e^+ e^-\to 3(\pi^+\pi^-)$ cross section measured by
      CMD3 detector (dots) in comparison with the BaBar data (open
      circles). Line shows the  $P\bar P$ threshold.
    }
    \label{cmd6pi}
  \end{center}
\end{figure}
%

\acknowledgements{%
We are grateful to VEPP2000 team for the excelent machine operation. 

This work is supported in part by FEDERAL TARGET PROGRAM "Scientific
  and scientific-pedagogical personnel of innovative Russia in 2009-2013"
  and by the grants  
RFBR 09-02-01019,  RFBR 09-02-00643, RFBR 09-02-00276,
RFBR 10-02-00253, RFBR 10-02-00695, RFBR 11-02-00112.

}


%

}  
